\documentclass[twocolumn,aps,prl,showpacs]{revtex4}
\usepackage{graphicx}
\usepackage{epsf}
\usepackage{amsmath}
\usepackage{amssymb}
\usepackage{color}
\begin{document}

\title{Topological classification of adiabatic processes}

\author{Dganit Meidan}
\affiliation{Dahlem Center for Complex Quantum Systems and Institut
f\"{u}r Theoretische Physik, Freie Universit\"{a}t Berlin, 14195
Berlin, Germany}

\author{Tobias Micklitz}
\affiliation{Dahlem Center for Complex Quantum Systems and Institut
f\"{u}r Theoretische Physik, Freie Universit\"{a}t Berlin, 14195
Berlin, Germany}

\author{Piet W.~Brouwer}
\affiliation{Dahlem Center for Complex Quantum Systems and Institut
f\"{u}r Theoretische Physik, Freie Universit\"{a}t Berlin, 14195
Berlin, Germany}

\date{\today} 


\begin{abstract}
Certain band insulators allow for the adiabatic pumping of quantized charge or spin for special time-dependences of the Hamiltonian. These "topological pumps" are closely related to two dimensional topological insulating phases of matter upon rolling the insulator up to a cylinder and threading it with a time dependent flux. In this article we extend the classification of topological pumps to the Wigner Dyson and chiral classes, coupled to multi-channel leads. 
The topological index distinguishing different topological classes is formulated in terms of 
the scattering matrix of the system. 
We argue that similar to topologically non-trivial insulators, topological pumps are characterized by the appearance of protected
gapless end states during the course of a pumping cycle. We show that this property allows for the pumping of quantized charge or spin in the weak coupling limit. Our results may also be applied to two dimensional topological insulators, where they give a physically transparent interpretation of the topologically non-trivial phases in terms of scattering matrices.

\end{abstract}
\maketitle

Topological insulating states of matter differ from regular band insulators by the fact that they  
support protected gapless surface states.  Holding promise for numerous applications, this observation has considerably motivated the search for materials that realize such topological phases. The first experimental observation of a topologically 
nontrivial insulator dates back to the discovery of the quantum Hall effect 30 years ago~\cite{KvKlitzingPRL1980,SMGirvin1999}. The recent discovery of the quantum spin Hall effect~\cite{CLKanePRL2005a,MKonigScience2007,MZHasanRMP2010}, has lead to a full classification of insulators with topological order based on their underlying symmetries and spatial dimensions~\cite{SRyuNJP2010}. 

The study of the quantum Hall effect has instigated numerous  theoretical works dedicated  to the understanding of its non-trivial topological nature. One particular appealing argument was introduced by Laughlin~\cite{RBLaughlinPRB1981}, who considered a pump formed by placing the two dimensional system on a cylinder and 
threading it with a time dependent magnetic flux. As the flux is varied periodically in time, 
an integer number of charges are transferred across the pump upon completing one cycle. 
This charge quantization is directly related to the quantized Hall 
conductance of the underlying Hall insulator~\cite{DJThoulessPRB1983,QNiuJPA1984,RTaoPRB1984,QNiuPRB1985,DJThoulessPRB1989,XQWenPRB1990,SHSimonPRB2000}.

In this communication we extend Laughlin's considerations 
to pumps formed by two dimensional insulators belonging to the Wigner Dyson and the 
chiral classes. Based on their underlying symmetries, applying the above construction imposes a symmetry constraint on the 
pumping cycle. This allows for the classification of topological pumps in terms of invariants of
their scattering matrix and gives rise to a physically transparent interpretation of the topologically non-trivial phases in terms of quantized pumping properties. 
Similarly to topologically non-trivial two-dimensional insulators, topologically non-trivial pumps 
are characterized by the appearance of 
gapless end states during the course of a pumping cycle~\cite{RBLaughlinPRB1981,LFuPRB2006,RRoyCondmat2011}.
We show that in the weak coupling limit the nontrivial pumps allow for noiseless 
pumping of quantized charge or spin. This article extends previous work 
by the authors on topological pumps with time reversal restriction connected to single channel leads ~\cite{DMeidanPRB2010}.

\begin{figure}
\begin{center}
\includegraphics[width=0.3\textwidth]{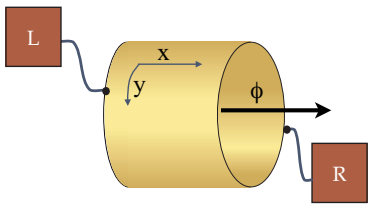}
\caption[0.5\textwidth]{ A pump formed by rolling a two dimensional insulator on a cylinder and threading it with a time dependent flux $\phi $. }
\label{pump}
\end{center}
\end{figure}

Consider a pump formed by placing  the two-dimensional insulator on a cylinder and threading 
it with a magnetic flux which is varied in time~\cite{DJThoulessPRB1983,QNiuJPA1984}, see 
Fig.~\ref{pump}. The cylinder is connected to ballistic leads with $N$ (or in presence of spin degeneracy $2N$) 
transverse channels, 
 and scattering off the pump 
can be described in terms of $2N\times2N$  (or in presence of spin degeneracy $4N\times4N$) 
scattering matrix. Assuming the pump's extension exceeds the attenuation length associated with 
the bulk energy gap, the scattering matrix reduces into two 
$N\times N$ ($ 2N\times 2N$) blocks describing the reflection of  electrons incident from the left and right leads.
The underlying symmetries $\hat{\cal{S}}$ of the two dimensional Hamiltonian $\hat{\cal{S}} \hat{H} \hat{\cal{S}}^{-1}=\pm  \hat{H}$, impose a similar
restriction on the pumping cycle, $\hat{\cal{S}} \hat{H}(t) \hat{\cal{S}}^{-1}= \pm \hat{H}({\cal S}^{-1}(t))$. In this paper we  study the topological classification of pumps formed of insulators belonging to the Wigner Dyson and chiral classes, characterized by the presence or absence of 
time-reversal $\hat{\cal{S}}=\Theta$, $\Theta H\Theta^{-1}=H $ and sublattice symmetries $\hat{\cal{S}}={\cal C}$, $CHC^{-1} = -H$ as well as the presence or absence of strong spin orbit interactions.

The scattering matrix of the pump is related to the two-dimensional Hamiltonian   through the general formula~\cite{CMahaux1969}
\begin{eqnarray}\label{sct_Ham}
S = 1 + 2i W^\dagger\left( H - i\pi W W^\dagger \right)^{-1} W,
\end{eqnarray}
 where W describes the coupling to the leads. Here we assume the leads  couple equally  to both spin orientations at the edges of the insulator 
and that the coupling itself does not break time reversal or sublattice symmetry.

This relation 
allows to obtain the symmetry restrictions on the reflection matrices, summarized in 
Table~I~\ref{classes}.
\begin{table}[h]
\label{classes}
\begin{tabular}{| c | c | c |c  c|| c |} \hline
class &$ \Theta $&${\cal C}$ &  \multicolumn{2}{|c||}{symmetry restriction} & index\\
\hline 
A &0&0 & \multicolumn{2}{|c||}{$r(t)\in U(2N)$ }& $\mathbb{Z}$  \\
AI & 1& 0& \multicolumn{2}{|c||}{$r(t)= r^T(-t)$ } & 0\\
AII & -1& 0& \multicolumn{2}{|c||}{$r(t)=\sigma_yr(-t)^T\sigma_y$}&$\mathbb{Z}_2 $\\
AIII &0 &1 & \multicolumn{2}{|c||}{$r(t)=r^\dag(t) $}&0\\
BDI & 1&1 & $r(t)=r^\dag(t) $& $r(t) = r^T(-t)  $&0\\
CII & -1& 1&$r(t)=r^\dag(t) $ & $ r(t)=\sigma_yr(-t)^T\sigma_y$&0\\
\hline
\end{tabular}
\caption{Symmetry restrictions on reflection matrices belonging to the Wigner Dyson (first three rows) and chiral 
(last three rows) classes.
$\Theta$ and ${\cal C}$ are the time reversal and sublattice symmetry operations, respectively.  
Last column: Classes which allow for 
nontrival topological $\mathbb{Z}$ or $\mathbb{Z}_2$ invariants.}
\end{table}

Similarly to topologically non-trivial two-dimensional insulators, topologically non-trivial pumps 
are characterized by the appearance of gapless end states during the course of a pumping cycle. 
These emerge as resonances of the scattering matrix that introduce 
a $\pi $  phase shift of the incoming scattering states. (The presence of a bound state 
at the edge of the sample follows from Eq.~\eqref{sct_Ham} or from 
a Bohr-Sommerfeld quantization argument, when the scattering 
states acquire a $ \pi$  phase shift at resonance.)
In a topologically non-trivial pump, the appearance of resonances \textit{during a pumping cycle} 
is topologically protected and is independent of smooth deformation of the Hamiltonian, although the 
specific moment of it's appearance may vary.  As we discuss below, these resonances manifest in the form 
of topologically protected pumping properties.

\begin{figure}
\begin{center}
\includegraphics[width=0.3\textwidth]{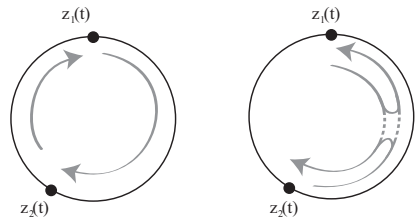}
\caption[0.5\textwidth]{A pumping cycle can be visualized as a set of trajectories formed by the ring coordinates, shown for $ N=2$. Left hand side portrays a non-trivial winding of the center of mass coordinate, while the right hand side shows avoided level crossing corresponding to a non-trivial winding of the center of mass.}
\label{center_of_mass}
\end{center}
\end{figure}

In order to relate the appearance of resonances during a pumping cycle to topological properties 
of the reflection matrix we note that the eigenvalues of the unitary reflection matrix are 
restricted to the unit circle, $\{z_1=e^{i\phi_1}, ... , z_N=e^{i\phi_N}\} $. The  pumping cycle can be 
visualized as the set of trajectories formed by these coordinates $\{z_1(t), ... , z_N(t)\} $ on the unit circle, 
such that the original set of eigenvalues is recovered after a cycle is completed. 
From Eq. \eqref{sct_Ham}, 
the appearance of an edge state, i.e., a scattering resonance,  corresponds to an eigenvalue $z_i=-1$ in the 
resonant channel $i$. As slight deformations of the Hamiltonian may shift the eigenvalues and 
 lead to detuning away from the resonance, topologically protected resonances can only arise 
if the trajectory of the eigenvalue forms a non-contractable loop around the unit circle. Moreover, 
in the absence of any symmetries, any crossings of energy levels (and similarly 
of eigenvalues $z_i$)  are generally avoided, or
can be  avoided in the presence of  small perturbations, see Fig. \ref{center_of_mass}. 
Hence, in the absence of any symmetries, topologically protected resonances can only arise 
 if the center of mass of the ring coordinates, $\Phi=\sum_{i=1}^N\phi_i$, forms a non contractable 
loop within a pumping cycle.  The sum of the eigenvalue phases is related to the determinant, 
$\textrm{det}\,r = \Pi_i z_i = e^{i\Phi} $ and the winding of the phase  during the course of a pumping cycle, 
gives rise to an integer  index $ n$:  
\begin{align}
n 
&=\oint_0^T {dt\over 2\pi} \, \dot{\Phi}
= \oint_0^T {dt\over 2\pi}  {d\over dt} \ln {\rm det}\, r .
\end{align}

Symmetry constraints on the pumping cycle restrict the `dynamics' of the eigenvalues during the cycle.
In particular, a time reversal 
restriction relates the time-evolution of $r(t)$ during the  second part of the cycle $T/2\leq t < T$ to the evolution 
during the first part $0\leq t< T/2$,
\begin{align}\label{TRS}
r(t) &= \hat{O} r^t(T-t)\hat{O}^{-1}.
\end{align} Here $\hat{O}=e^{i\pi \hat{S}_y/\hbar}$ is the unit matrix for spinless electrons and the  
Pauli matrix $\sigma_y$ for spinfull  electrons. 
As a consequence, any winding of the phase $\Phi $ during the first part of the cycle is undone in the course of 
completing the cycle, $\Phi(t) =\Phi(T-t)$, 
resulting in 
  \begin{align}
\nonumber
n
&=\int_0^{T/2} {dt\over 2\pi} \left( \dot{\Phi}(t) + \dot{\Phi}(T/2-t) \right) =0,
\end{align} 
which expresses the fact pumps with time reversal constraint cannot pump charge.

\begin{figure}[b!]
\begin{center}
\includegraphics[width=0.40\textwidth]{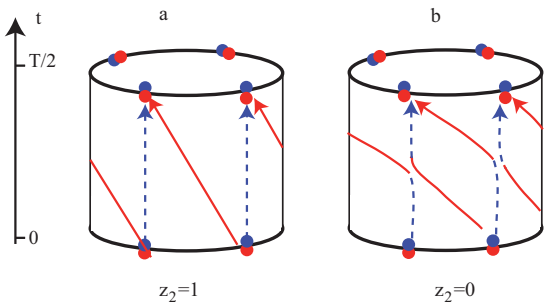}
\caption[0.5\textwidth]{ Fig. a)  shows a non-trivial winding of the relative phase acquired between a Kramer's pair after half a cycle corresponding to the exchange of partners. The higher winding of the relative phase shown in Fig. b) involves  additional crossing points away from the TRIM which are generally avoided, leaving no net phase difference between the Kramers pairs.}
\label{relative}
\end{center}
\end{figure}

The time reversal constraint allows, however, for a more subtle classification,  
when the spin rotational symmetry is broken as a result of strong spin orbit scattering (e.g. in the  symplectic class AII)~\cite{LFuPRB2006,DMeidanPRB2010}.
The time reversal restriction~\eqref{TRS} in combination with the periodicity of the pump ensures the 
existence of two time reversal invariant moments (TRIM) $ t_i = 0,T/2$ throughout the pumping cycle. 
At these moments, the reflection matrix is time-reversal symmetric, and its eigenvalues form 
Kramers degenerate pairs. While the time reversal restriction~\eqref{TRS}  inhibits any winding of $\Phi $, 
topologically protected resonances may still arise if the sum of the  phase differences  acquired 
by the Kramer pairs during half a cycle has a topologically non-trivial winding. As we shall see below, the latter is defined up to multiples of $4\pi $, giving rise to a $ \mathbb{Z}_2$ index instead of a $\mathbb{Z} $ index (A similar argument is made in Ref. \onlinecite{LFuPRB2006} with respect to the "time-reversal polarization").

At the time reversal invariant moments, the eigenvalues of the reflection matrix occur in pairs, $(e^{i\phi_n}, e^{i\bar{\phi}_n})$ with $n=1,....,N$. 
This introduces two possibilities. After half a cycle of the pump is completed, the pairs can either recombine,
$\phi_n(T/2)=\bar{\phi}_{n}(T/2)$ 
or exchange partners $\phi_n(T/2)=\bar{\phi}_{n-1}(T/2)$. In the latter scenario, 
the sum of all  phase differences  acquired between former Kramers pairs, can pick up a phase of multiple of $2\pi $ 
 when evolving into the two-fold degenerate configuration at $t=T/2$, see Fig.~\ref{relative} a.
We note that the crossing of the eigenvalues at the TRIM is protected by time reversal symmetry and 
therefore cannot be lifted by small deformations of the Hamiltonian. 
Moreover, higher winding of the 
relative phase, $ \delta\varphi = (\varphi -\bar{\varphi })$, 
corresponding to the Kramers pairs recombining with next to nearest neighbors 
eigenvalues, inevitably involve  additional crossing points away from the TRIM. 
 Here  $\varphi= \sum_{i=1}^N\phi_i $ and $ \bar{\varphi}=\sum_{i=1}^N\bar{\phi}_i$. 
As these are not 
protected by time reversal symmetry they  are generically avoided or can be removed by small deformations of the Hamiltonian, see Fig.~\ref{relative} b.
Hence,  an odd number of  winding of the relative phase 
 by $2\pi $ 
 is topologically protected, while an even number of winding is topologically trivial. 
 The relative phase $\delta\varphi$ acquired during half a cycle 
 can be expressed in terms of the products of eigenvalues $Z = e^{i\varphi} $ 
 and   $\bar{Z} =e^{i\bar{\varphi}}$, as 
 \begin{align}
 \nonumber
 \theta &= \int_0^{T/2} dt\, \delta \dot{\varphi} 
 = \ln\frac{Z(T/2)\bar{Z}(0)}{Z(0)\bar{Z}(T/2)} 
 \end{align}
The presence of even or odd winding number of the relative phase  
is then expressed in terms of the $\mathbb{Z}_2$ index
\begin{align}
\label{z2}
 e^{i\theta/2} = \frac{\sqrt{Z(T/2)\bar{Z}(0)}}{\sqrt{Z(0)\bar{Z}(T/2)}},
\end{align}
which takes the values $e^{i\theta/2}=\pm 1 $. Using the relation between $Z$ and the Pfaffian, $Z =\textrm{Pf}(i\, r\, \sigma_y)$,
this result   can be formulated in terms  of the reflection matrix:
\begin{align}
\label{z2_r}
e^{i\theta/2}= \frac{\textrm{Pf}(i\,r(T/2)\sigma_y)}{\textrm{Pf}(i\, r(0)\sigma_y)}
\frac{\sqrt{\textrm{det}r(0)}}{\sqrt{\textrm{det}r(T/2)}}.
\end{align}
where the same branch of the square root is chosen in the numerator and denominator. 
Eq.~\eqref{z2_r} is the $N$-channel generalization of the $\mathbb{Z}_2$-index in Ref.~\onlinecite{DMeidanPRB2010}. The same result appeared in Ref. \onlinecite{ICFulgaCondMat2011} in the context of a classification of $2D $ topological insulators.

Finally, we consider a system with chiral, i.e. sublattice symmetry. 
Contrary to time-reversal symmetry, the sublattice symmetry is
present at every instant in the cycle, 
\begin{align}
r(t)=r^\dag(t).
\end{align}  
As the reflection matrix is hermitian at all times, its eigenvalues are restricted  to the 
real axis, i.e. $z_i=\pm1$  throughout the cycle which 
prevents a non-trivial topology of the trajectory $r(t) $ during a cycle. 
Consequently all pumps with a chiral symmetry are  topologically trivial.

Following the discussion above, we may classify the pumps belonging to the Wigner Dyson and the chiral classes 
based  on the presence of absence of time reversal, chiral and spin rotational symmetry. The result is 
summarized in the last column of Table I. 
The classification based on the reflection matrix of the one dimensional pump reproduces the  corresponding table for two-dimensional insulators in the Wigner Dyson and chiral classes.

The appearance of protected gapless edge states during the course of a pumping cycle alters 
the pumping properties. In contrast to their trivial counterparts, topological nontrivial pumps 
allow for the pumping of quantized charge or spin. In the absence of time reversal or chiral symmetries 
(class A),  the quantization of the charge is evident as the charge pumped through the insulator \cite{MButtikerZPB1994,PWBrouwerPRB1998} is proportional 
to the topological index itself~\cite{DJThoulessPRB1983,QNiuJPA1984,ILAleinerPRL1998,SHSimonPRB2000,AAndreevPRL2000,YMakhlinPRL2001,JEAvronPRL2001,OEWohlmanPRB2002,MMoskaletsPRB2007}:
\begin{align}
Q = 
& {e \over 2\pi i} \oint dt\,  \text{tr}  \left( {d\,\hat{r} \over dt} \hat{r}^\dag \right) = e n,
\end{align} where the trace is taken over the $N$ channels. 

Imposing a time reversal restriction on the pumping cycle inhibits the pumping of charge by restricting  the winding of the phase of $\textrm{det}\, r$.
Nonetheless, topological pumps with a
 time reversal restriction allow for the pumping of quantized spin 
 even in the presence of spin-orbit scattering. 
 We notice that contrary to topological charge pumps,  the spin $\vec{S}$ pumped in a cycle
  \begin{align}
\label{s}
\vec{S}
&= {\hbar \over 2\pi i} \oint dt\, \text{tr} \left( {d \hat{r}_\alpha\over dt}
 \hat{r}_\alpha^\dagger \vec{\sigma} \right)
\end{align} 
   is not directly related to the $\mathbb{Z}_2$ index \eqref{z2_r}. Hence its quantization is in general 
 not ascertained. 
Instead, quantization becomes asymptotically exact in the weak coupling limit when the 
 coupling to the lead becomes arbitrary small~\cite{CommetWeakCoupling}. 
 
 Following the above discussion, a topological spin pump (class AII) crosses an odd 
number of resonance pairs during the course of a cycle.  As multiple winding of the phase can be removed by small perturbations, 
a non-trivial pump generically crosses a single resonance 
pair during a cycle. In the weak coupling limit, the $N$ channels decouple and only the resonant channel has 
 a time-dependent phase shift. 
Since only one channel contributes to the pumped spin, the considerations of Ref.~\onlinecite{DMeidanPRB2010} for the single 
channel case, $N=1$, can be applied: 
 The typical time scale at which 
 the resonance at time $t_i $ is traversed  vanishes in the weak coupling limit, as the the broadening of the 
 energy levels  goes to zero. Contrarily, the time scale on which the spin quantization axis,  $\vec{e}_\phi(t)$,  changes 
 depends exclusively on microscopically details of the insulator and is independent of 
the coupling to the leads. 
Therefore, in the weak coupling limit, 
the reflection matrix of the resonant level describes a   
rotation around a fixed axis, and quantization
$\vec{S} = \hbar \vec{e}_\phi(t_i)$ readily follows~\cite{DMeidanPRB2010}.

These results  can be extended to systems in which the gap arises due to electron electron interactions. 
The main difference arises as the many body nature of an interaction induced gap can change the Fermi sea of non-interacting electrons 
into $d$  many-particle ground states. 
As a consequence, ground states may be interchanged in the course of 
a pumping cycle, and the pump may operate with an extended period, which is a multiple of the period $T$
at which the Hamiltonian is varied
 see discussion in~\cite{SHSimonPRB2000,DMeidanCondMat2011}. In the weak coupling limit, depending on the symmetry group of the Hamiltonian, each extended periodicity can be characterized by a $\mathbb{Z} $ (class A) or $ \mathbb{Z}_2$ (class AII) topological index, corresponding 
for a transfer of an integer charge $n$ or spin $\hbar$ during the \textit{extended} cycle, respectively. Consequently, a topological pump with an extended pumping cycle, $ q T$, due to interactions,  transfers on average a fractional charge $ en/q$  or spin $ \hbar/q$ during a cycle $T$~\cite{SHSimonPRB2000,DMeidanCondMat2011}. 
 
In conclusion, 
we have extended Laughlin's construction of  pumps formed by two dimensional insulators to the Wigner Dyson and chiral classes, coupled to multi-channel leads. This mapping allows for a  pedestrian derivation of the topological classification of insulators in terms of the reflection matrices of the corresponding pumps. We provide a  physically transparent interpretation of the topologically non-trivial phases based on their quantized pumping properties.

Upon finishing this work we became aware of a related work in Ref.~\onlinecite{ICFulgaCondMat2011}.
This work is supported by the Alexander von Humboldt Foundation in the framework of the Alexander von Humboldt Professorship, endowed by the Federal Ministry of Education and Research.


\begin{thebibliography}{10}
\bibitem{CommetWeakCoupling}
We note that the weak coupling limit $\Gamma \rightarrow 0 $ is restricted by
  the pumping period $1/\Gamma<T$ to allow for relaxation in the leads.
\end{thebibliography}
\end{document}